\documentclass[12pt]{article}
\usepackage{epsf}
\usepackage{amsmath}
\usepackage{graphics}
\usepackage{cite}

\renewcommand{\thefootnote}{\#\arabic{footnote}}
\begin{document}

\newcommand{\gtrsim}{ \mathop{}_{\textstyle \sim}^{\textstyle >} }
\newcommand{\lesssim}{ \mathop{}_{\textstyle \sim}^{\textstyle <} }

\newcommand{\rem}[1]{{\bf #1}}

\renewcommand{\thefootnote}{\fnsymbol{footnote}}
\setcounter{footnote}{0}
\begin{titlepage}

\def\thefootnote{\fnsymbol{footnote}}

\begin{center}
\hfill gr-qc/yymmnnn\\

\vskip .5in
\bigskip
\bigskip
\Large 

{\bf Comment on ``Can Black Holes be Torn Up by a Phantom
in Cyclic Cosmology?" by X. Zhang. {\tt arXiv:0708.1408 [gr-qc]}}

\vskip .45in

{\bf Paul H. Frampton\footnote{frampton@physics.unc.edu}} 

\normalsize

{\em Department of Physics and Astronomy, University of North Carolina,}

{\em Chapel Hill, NC 27599-3255. }

\end{center}

\vskip .4in

\begin{abstract}
In a recently archived paper by Zhang\cite{Zhang}, it is claimed that
before turnaround in a cyclic model
two unexpected events happen: (1)
black holes cease to contract and begin to expand;
(2) separated causal patches start to reconnect.
We show that both conclusions
are erroneous and result from the author's choice of variables.
\end{abstract}

\end{titlepage}

\renewcommand{\thepage}{\arabic{page}}
\setcounter{page}{1}
\renewcommand{\thefootnote}{\#\arabic{footnote}}

\newpage

\newpage

The article by Zhang\cite{Zhang} addresses the question of the turnaround
in an expanding universe characterized by the Friedmann
equations
\begin{equation}
\left( \frac{\dot{a}}{a} \right)^2 = 
\frac{8 \pi G}{3} \rho \left( 1 - \frac{\rho}{\rho_C} \right)
\label{Friedmann}
\end{equation}
\begin{equation}
\frac{\ddot{a}}{a} = -\frac{1}{6}
\left\{ \rho \left( 1 - \frac{\rho}{\rho_c} \right)
+ 3 \left[ p \left( 1 - \frac{2 \rho}{\rho_C}
\right) - \frac{\rho^2}{\rho_C} \right] \right\}.
\label{Friedmann2}
\end{equation}
The author introduces the unphysical variables:
\begin{equation}
\rho_{eff} = \rho \left( 1 - \frac{\rho}{\rho_C} \right)
\label{rhoeff}
\end{equation}
\begin{equation}
p_{eff} = p \left( 1 - \frac{2\rho}{\rho_C} \right) - \frac{\rho^2}{\rho_C}
\label{peff}
\end{equation}
and defines $x=\rho/\rho_C$. In terms of the real density and pressure one 
easily finds:
\begin{equation}
\rho_{eff} + p_{eff} \equiv (\rho + p) (1 - 2x)
\label{unphysical}
\end{equation}
For back holes of mass $M$, area $A$,  Babichev, {\it et al}
\cite{BDE} correctly found that
\begin{equation}
\dot{M} = 4 \pi A M^2 (\rho + p)
\label{BDE}
\end{equation}
which shows that for $p/\rho < -1$, the black hole mass always decreases, $\dot{M} < 0$.

\bigskip

\noindent By contrast, Zhang chooses to rewrite Eq.(\ref{BDE}) in terms
of his unphysical variables as
\begin{equation}
\dot{M} = 4 \pi A M^2 (\rho_{eff} + p_{eff}) =
4 \pi A M^2 (p + \rho) (1 - 2x).
\label{Zhang}
\end{equation}
and concludes erroneously that when $x = 1/2$ 
the sign of $\dot{M}$ changes.
A second erroneous claim that separate causal patches
start to reconnect when $x=1/2$, in contradiction to the discussion in 
\cite{BF}, arises from the same nonjudicious choice of variables.

\begin{center}

{\bf Acknowledgements}

\end{center}

\bigskip
This work was supported in part by the
U.S. Department of Energy under Grant No. DE-FG02-06ER41418.

\end{document}